\documentstyle[12pt]{article}
\newcommand{\be}{\begin{eqnarray}}
\newcommand{\ee}{\end{eqnarray}}

\begin{document}

\begin{tabbing}
\`SUNY-NTG-94-14\\
\`March. 1994
\end{tabbing}
\vbox to  0.8in{}
%double spacing:
%\vbox to  0.4in{}
\centerline{\Large \bf Where the excess photons and dileptons  }
\centerline{\Large \bf in SPS nuclear collisions come from?}
\vskip 2.5cm
%double spacing:
%\vskip 1.25cm
\centerline{\large  E. V. Shuryak and L. Xiong }
\vskip .3cm
\centerline{Department of Physics}
\centerline{State University of New York at
Stony Brook}
\centerline{Stony Brook, New York 11794}
\vskip 0.35in
\centerline{\bf Abstract}
\indent
  Recently the first  single photon spectra from CERN energy
heavy-ion collisions
were reported by WA80, while NA34/3 and NA38 have obtained the spectra
for dileptons  with  the mass up to 4-5 GeV.
The production rates for photons and dileptons significantly increase
when reactions involving the  $A_1$ meson  are included.
However, with the conventional expansion scenario,
the absolute yields are still significantly smaller than
the observed ones. It may indicate that expansion in
the ``mixed state" takes much more time.

\eject

   Observation of the ``penetrating probes", photons and dileptons, was for
a long time considered
to be the best ways to get information about the
early hot stage
of nuclear collisions  \cite{Shuryak_78}, see recent reviews in \cite{RK}.
 Unfortunately, such experiments are difficult,
% challenging
%compared to more standard measurements of hadronic spectra,
 so
only recently their first preliminary results
were reported.

 In this letter we discuss whether one can or cannot explain these
observations.
%, just by combining
% available estimates of the production rates with
%some popular space-time picture of the collisions.
The first general question to address is {\it at what stage of
the collision} the observed photons and dileptons are created.
 Three qualitatively different stages  are generally considered
 (i) the {\it quark-gluon plasma} (QGP) phase; (ii) the {\it mixed} stage; and
(iii) {\it hadronic} stage. Depending on which one is considered,
the production rates is then evaluated  differently.

  At RHIC and LHC energies
a production of ``hot glue" with $T\sim 2-3 T_c$ may provide a
 window of opportunity
to see the QGP signals \cite{hotglue}, but at SPS
  the  plasma neither can be
{\it significantly hotter} than $T_c$, nor
it may occupy a significant
 space-time volume: so it will be neglected in this paper.

   Speaking about
 the mixed stage, we use it in
a broad
 sense, as a narrow  temperature interval $\Delta T <<
T_c$ in which the pressure remains nearly constant, but
the energy density changes a lot\footnote{
Whether the actual phase transition is first or second order
in mathematical sense, or it is just a ``rapid crossover", the existence
of the mixed phase
 in this broader sense is supported by all
available lattice data on QCD thermodynamics.
}
from
 $\epsilon_{min}$ to $\epsilon_{max}$, different by about
an order of magnitude.
   With such broad definition,
the {\it mixed phase} clearly
 dominates the space-time evolution of the system
at
CERN SPS energies.
  Although this fact makes  theoretical predictions difficult,
 a proximity to the phase transition
 makes the experimental information more challenging and interesting.

   Photon production rate in hot hadronic matter was first
considered in a pion gas, but then
  the $\rho$ mesons where shown to be important
due to
 $\pi\rho\rightarrow\pi\gamma$ \cite{Kapusta_etal}.
Further studies \cite{XSB} have found that
the  reaction
 $\pi\rho\rightarrow A_1 \rightarrow \pi\gamma$ increase
 the rate by another factor 3 \footnote{
Additional small enhancement comes from constructive
 interference of the $A_1$-based and
other diagrams \cite{Song}.}.
As a result, in the relevant kinematic region
  the hadronic component of the mixed phase ``overshines" the
plasma one, even as a rate (per space-time volume).

  Let us start with as simple and conventional
model of the space-time evolution
  as possible. We called it the
 {\it the frozen motion model} (FMM),
 assuming that: (i) rapidity of any matter element
is unchanged during the expansion;
 (ii)
 there is no transverse expansion;
 (iii) at the {\it end} of the  mixed phase,
$\epsilon=\epsilon_{min}$ the pion density corresponds
 to that of the ideal pion gas; (iv) this moment can be obtained
for each rapidity interval
from the {\it experimental} pion rapidity distribution\footnote{
Only at this last point we deviate from
 the usual scaling (Bjorken) picture, which additionally
demands a {\it  plateau} $dN_\pi(y)/dy=const$. Thus, FMM is its minimal
 extension to not-so-high SPS energy, at which
 the  end line of the mixed phase (or the
freeze-out line) are definitely {\it not}
hyperbolae in the z-t plane.
}
$dN_\pi(y)/dy$. If so,
 the  longitudinal size of a matter element is $\Delta z=\tau \Delta y$
 where $\tau$ is the
 proper time  and $\Delta y$ is its rapidity spread .
As a result,
 the space-time volume spent in the mixed phase by such  matter element  is
\be
\Delta V_4= \Delta y {V_\pi^2 \over 2 A_t}  ({dN_\pi(y) \over dy})^2
\ee
where $A_t$ is the (projectile)
transverse area,  and $V_\pi$
is the volume per pion
at the {\it end} of the mixed phase (it depends on the assumed $T_c$  and,
the upper end of the
possible range $T_c=180 MeV$ corresponds to $V_\pi=3.7 fm^3$).
For example, an element
in the matter central rapidity ($y_c\approx 2.6$ for S-W collisions)
will stay in the mixed phase for the proper time
$\tau\approx 10 fm/c$,
\footnote{
One should not confuse this number neither with
the {\it freeze-out} time (which is about twice larger), nor with
 the
{\it pion emission} time
 as measured by the pion interferometry.
The latter depends on the {\it curvature} of the decoupling
surface, and  in general can
be different, for example very short, as suggested by available data.}
which is about the same as given
by most event generators.

  The first  statistically significant data on
 the {\it single} photon  production
 were recently reported
 \cite{WA80} \footnote{
Statistically {\it insignificant} excess of photons ($7.5 \pm 11 \%$) was also
reported by CERES \cite{ceres}.
%, and this excess (if any) seems to be seen at
%different place, at $p_t\sim 2-3 GeV/c$.
}.
 These data are shown
for central events  (7.7\% of the minimum bias cross section)
in Fig.1 as a $p_t$ spectrum.
A  striking fact is the ``apparent temperature"
(a  $p_t$ slope) is about 3/4 of that for the pions:
%  It is well known by now, that the observed pion $p_t$ spectrum has a
%slope which is a combination of the final (or break-up)
%temperature $T_f$ and transverse collective motion
%\footnote{The
%exact magnitude of both components is still unknown. At AGS a detailed
%information
%  from HBT interferometry was recently supplemented by deutron
%production \cite{BM}. Both point toward relatively low
% break-up temperature
%$T_f\approx 140 MeV$ and mean transverse expansion
%velocity $v_t\sim 0.3$.}.
 it fits well the idea  that  these photons are
emitted {\it before} the transverse
expansion takes place.

 Using the rates from \cite{XSB}
and the FMM described above, we get
  the absolute photon yield at central rapidities.
  The results, shown in Fig.1., demonstrate that the
 $T_c= 180$ MeV case fits the slope, but not the magnitude!
 Which
 elements of the  calculation could then be modified, in order
to increase the photon yield?

One may question whether at $T\approx T_c$ the masses and widths
of $\rho,A_1$ are still at their {\it vacuum} values, or rather,
due to chiral symmetry restoration, one may expect
$m_{A1}(T\rightarrow T_c) \rightarrow m_\rho(T\rightarrow T_c)$.
However, we have found that the photon yield happens to depend
weakly on the $A_1$ mass:
 a gain from smaller Boltzmann
suppression is compensated by a shift away from the
 top of the resonance. Similarly, we have found weak  sensitivity to the $A_1$
width, and conclude that  it is impossible
 to increase the rate by $A_1$  modifications alone.

 One may now question the conventional FMM picture
 of space-time evolution.
As  pointed out long ago
 \cite{SZ_vanHove}, the matter in the mixed phase is very
{\it soft}: its pressure is
very low compared to its high energy density.
Therefore, one may look at
an unconventional scenario for the
AGS-SPS energies. Thus we propose a {\it slow expansion
model }\footnote{The hydrodynamics of fireball expansion
in the corresponding region of initial conditions currently
under investigations \cite{HS} shows that for some particular initial
conditions it is indeed the case.}
 (SEM), in which the observed pion rapidity distribution is formed
{\it after} the mixed phase is over.
If the expansion  of the mixed phase is slower than assumed above,
more photons can be produced. For example, if
a mixed phase  expands in
$\tau \sim 30-40 fm/c$ (instead of 10 fm/c in FMM), one can
 naturally explain the WA80
 data. A simple  test for this proposal is provided by the following: an
 excess photons  should all be peaked {\it  around the  central rapidity}
 $y\approx y_c$.

   Let us now turn to
dilepton production. Two muon experiments, NA34
\cite{NA34} and  NA38
\cite{NA38} have reported the  ``unaccounted"
excess of dileptons  in central S-W nuclear collisions,
 compared to pA and  peripheral collisions.
The excess is observed for pairs in the invariant mass region
$M\sim 1-2 GeV$,
 right where the contribution from the mixed phase is expected.

Among the dilepton production mechanisms, the fundamental
$\bar q q$ annihilation in the QGP phase
\cite{Shuryak_78} should  be significant for higher masses (and energies).
In the hadronic and mixed phase,  the $\pi^+\pi^-$ annihilation is
the dominant for $M<1$ GeV, then
the $K^+K^-$ annihilation becomes important also.
The dilepton production   for several processes in hot mesonic gas
were calculated  in \cite{Gale_Lichard}, to which
other important reactions involving
the $A_1$ meson were added recently \cite{Ko_etal}.
 We have recalculated the rate for the above processes
such as $\pi+ A_1 \rightarrow l^+l^-$ with the
somewhat different $\pi A_1\gamma$
interaction \cite{XSB} : the results generally agree:
the remaining ambiguity is mainly
due  to uncertain formfactors
\footnote{ The   form factors for $\pi\pi, KK$ annihilation
are both experimentally determined
\cite{form} for $ M < 2 $  GeV.
But it is not known for
 $pi A_1$ annihilation. In particular, its threshold
is right below the $\rho^\prime(1420)$ resonance,
whose contribution to
the $\pi A_1\gamma$ vertex remains unknown.
Following \cite{Gale_Lichard,Ko_etal} it is assumed to be the same as the pion
form factor.
%However,  because $\rho^\prime$ decays to
%$4\pi$ rather than $2\pi$, it is not very convincing.
}.

We have calculated the thermal dilepton spectra in the FMM model using
$T_c= 180$ MeV, see fig.2.
The comparisons with the experiments require the
acceptance of dileptons in certain rapidity window $[y1, y2]$, which is
\begin{equation}
	A( y, M) = \int _{ max[ 0, {tanh(y-y1)\over \sqrt{1-4m_l^2/M^2}}]}
^{ min[ 1, {tanh(y2-y)\over \sqrt{1-4m_l^2/M^2}}]}
f( \cos\theta ) d\cos\theta
\end{equation}
for the pair with invariant mass $M$ and rapidity $y$. $f(x)$ is the
 process dependent angular distribution  and is normalized within
$x={0, 1}$.
 The data reported by NA34 and NA38
lack the absolute normalization, therefore we have
 used the
Drell-Yan process as the benchmark.
We demand that our calculated curve
\footnote{ We have scaled the pp collision by
$(AB^{1/3} +A^{1/3}B)/2+(A^{2/3}-B^{2/3})^2/4
\ln(( 1-(A/B)^{1/3})/ (1+(A/B)^{1/3})) $ for central AB collisions.
Also we have included the K-factor as 2. }
goes through the largest mass data points, where
DY should definitely dominate.
(Our DY evaluation agrees well with
the those indicated in the experimental papers.)

The predicted dilepton spectra for
 three CERN experiments are shown in Fig.2.
% Note the predicted shoulders at thresholds,
%as well as some $\rho'$ trace.
Like for the photons,
a conventional FMM picture {\it does not} produce a
sufficient yield of dileptons
to explain  the data. The unconventional SEM can then be the explanation:
about the same factor is missing here
\footnote{Let us mention two more puzzles, also coming from
recent experiments. Brookhaven experiment E814 has observed a low-$p_t$
component in the $K^+,K^-$ spectra, and E802 has found
about 5 times more $\Xi$ hyperons than the conventional estimates.
The slow expansion scenario at mixed phase may help to explain all of them!}.

In order to test it
experimentally, one may look at two different things. The first,
as for photons, is the unusual rapidity distribution: in SEM one expects
larger dilepton excess at
central rapidities (NA45) compared to the forward ones (NA34, Na38).
The second,
 specific for dileptons, is a search for the
 possible ``shoulders" in the M-spectra.
Of particular interest is the $\pi A_1$ threshold:
it was  recently suggested \cite{dcc} that if a pion can come
from the hypothetical ``disoriented chiral condensate", it possesses
very small momentum and  then the peak develops near the $A_1$ mass!

  In conclusion, we have evaluated photon and dilepton yields
coming from the ``mixed phase" created in CERN SPS experiments.
 We have found, that even taking
parameters
at their extreme (e.g. take $T_c$ at its highest end, etc), one cannot
explain the observed photon and dilepton excess in the conventional model
which assumes constant velocity of the matter during the expansion.
Another proposed scenario, in which the mixed phase fireball expands
much slower, during the time
of about 30-40 fm/c, can definitely explain them all, and possibly
 some
other puzzles. Simple tests for this exciting scenario are proposed.

{\bf Acknowledgements}

We acknowledge interesting discussions of the subject with G.Brown,
P.Braun-Munziger and V.Koch.
This work is supported in part by the US Department
of Energy under Grant No. DE-FG02-88ER40388 and
No. DE-FG02-93ER40768.

\vfill\eject
\newpage
%\bibliographystyle{try}
%\bibliography{ref}

\begin{thebibliography}{10}

\bibitem{Shuryak_78}
E.Shuryak  Phys.Lett.78B:150,1978, Sov.J.Nucl.Phys.28:408,1978.

\bibitem{RK}P.V.Ruuskanen, Nucl.Phys. A544(1994) 169c.
J.Kapusta, Nucl.Phys. A566(1994) 45c.

\bibitem{hotglue} E.~Shuryak, Phys.Rev.Lett. {\bf 68} (1992) 3270;
 E.~Shuryak and Li Xiong, Phys.Rev.Lett., {\bf 70}
(1993) 2241.

\bibitem{SZ_vanHove} E.Shuryak and O.V.Zhirov, Phys.Lett.89B (1979) 253-255;
L.van Hove, Z.Phys.C21 (1983) 93.

%\bibitem{hydro}S.Chakrabarty, J.Alam,D.K.Srivastava, B.Sinha,
%Phys.Rev.D46 (1992) 3802.

\bibitem{WA80}R.Santo,  Nucl.Phys. A566(1994) 61c.
Single Photon and Neutral Meson Production from WA80, IKP-MS-93/0701, Munster
1993.
\bibitem{ceres}A.Drees, Nucl.Phys. A566(1994) 87c.
%\bibitem{E814} J.Stachel, Nucl.Phys. A566(1994) 183c.
\bibitem{Kapusta_etal}
J.~Kapusta, P.~Lichard, and D.~Seibert,
\newblock {\em Phys. Rev.}, {\bf D44} (1991)2774.

\bibitem{XSB}Li Xiong, E.Shuryak, G.E.Brown, Phys.Rev.D46 (1992) 3798.

\bibitem{Song} C.Song, Phys.Rev.C, 47 (1993) 2861.
\bibitem{HS} C.M.Hung and E.Shuryak, Hydrodynamics near the QCD phase
transition, SUNY-NTG preprint, 1994.

\bibitem{NA34}A.Mazzoni,Nucl.Phys. A566(1994) 95c
\bibitem{NA38}C.Lourenco,Nucl.Phys. A566(1994) 77c

\bibitem{Gale_Lichard} C. Gale and P. Lichard, Lepton pairs from
thermal mesons, McGill/93-8, TPI-MINN-93/19-T.
\bibitem{Ko_etal}C.Song, Che Ming Ko and C.Gale, Role of the $a_1$
meson in dilepton production from hot hadronic matter. Texas A\&M University
preprint.

\bibitem{form} L. M. Barkov etal, Nucl. Phys. B256 (1985) 365;
D. Bisello etal, Phys. Lett. B220 (1989) 321;
Ivanov et al. Phys. Lett. B107 (1981) 297.

\bibitem{dcc} Z.Huang, M.Suzuki and X.-N.Wang, preprint LBL-35337, 1994.

\end{thebibliography}

\newpage
\centerline{Figure Captions}

1. Transverse momentum spectrum of
single photons in 200 AGeV S-Au collisions  \cite{WA80}
for central events  (7.7\% of the minimum bias cross section).
The solid line stands for FMM and
$ T_c=180 $ MeV (two dotted lines correspond to
other choices $ T_c= 140, 220 $ MeV).
The individual
contribution from reactions
$\pi\pi\rightarrow \rho\gamma$ is shown by short dashed line, and from
$\pi\rho\rightarrow A_1 \rightarrow \pi\gamma $ by the dash-dotted one.

2.Mass spectra of dileptons production evaluated in the FMM model
($T_c= 180$ MeV) and integrated over the rapidity windows of three CERN
experiments are shown by
the solid curves.
The contributions of
 $\pi\pi,K^+K^-,\pi A_1$ annihilation are shown by  dotted,  dash-dotted and
 short-dash curves, while the Drell-Yan by the long-dash one.
 The data points are those
 reported by NA34 and NA38 (We simply took one of
the four available sets, 91 SU Psi, others are similar.)

\end{document}